\documentclass[reprint,amsmath,amssymb,aps,floatfix]{revtex4-2}
\usepackage{graphicx} 
\usepackage{dcolumn} 
\usepackage{bm} 
\usepackage{mhchem} 
\usepackage{textcomp} 
\usepackage{color} 
\usepackage{textcomp}
\usepackage{xr}
\usepackage{float} 
\usepackage[colorlinks=true,linkcolor=blue,citecolor=blue,urlcolor=blue]{hyperref} 
\begin{document}

\preprint{APS/123-QED}

\title{\mbox Polarity-dependent Electroadhesion at Silicon Interfaces with Nanoscale Roughness}

\author{Liang Peng$^{1}$\footnotemark[1]{*}}
\author{Stefan Kooij$^{1}$} 
\author{H. Tun\c{c} \c{C}ift\c{c}i$^{2}$}  
\author{Daniel Bonn$^{1}$} 
\author{Bart Weber$^{1,2}$} 
\affiliation{$^{1}$ Van der Waals-Zeeman Institute, Institute of Physics, University of Amsterdam, Science Park 904, 1098 XH Amsterdam, The Netherlands \\
$^{2}$ Advanced Research Center for Nanolithography (ARCNL), Science Park 106, 1098 XG Amsterdam, The Netherlands}

\begin{abstract}
We measure and model electroadhesion across multi-asperity silicon interfaces with nanometer scale roughness. When electrically biased, our naturally oxidized silicon interfaces display a leakage current consistent with the Fowler-Nordheim model and electroadhesion that can be successfully captured using a boundary element contact model. We show that polysilicon exhibits electroadhesion only under positive bias applied to the substrate monocrystalline wafer, which we interpret as a result of the reduced mobility of holes, with respect to electrons, within polysilicon. Overall, our findings reveal that electrostatic interactions can significantly enhance adhesion and friction between stiff and smooth surfaces, which can be important for example in precision positioning applications.
\end{abstract}

\maketitle

Adhesion plays a critical role in many practical applications, ranging from microelectromechanical systems (MEMS) \cite{zhao2003mechanics} and precision positioning \cite{niklaus2003method} in the semiconductor industry to haptic \cite{prahlad2008electroadhesive} and biomedical devices \cite{khalili2015review} in robotics. Numerous nanoscale studies have investigated the adhesion mechanisms ranging from van der Waals interactions \cite{lessel2013impact}, electrostatic interactions \cite{park2006,greenwood2023,greenwood2023,ayyildiz2018}, capillary forces \cite{asay2006effects,cassin2023nucleation,riedo2002kinetics} and covalent or hydrogen bonding \cite{li2011frictional,schall2021molecular,li2014effects, chen2006velocity,erbas2012viscous} at single-asperity contact interfaces. At larger, multi-asperity interfaces, the quantification of adhesion remains a major challenge. Recent experiments have advanced our understanding of van der Waals adhesion \cite{delrio2005role}, capillary adhesion \cite{peng2022} and covalent adhesion \cite{peng2023} at multi-contact interfaces. In contrast, few experiments address the underlying mechanisms of electroadhesion at macroscopic interfaces \cite{guo2015investigation,guo2019electroadhesion}, particularly those with nanoscale roughness that are industrially relevant—for instance, in precision positioning systems where unwanted electroadhesion may arise from tribocharging \cite{grosjean2023single,liao2024charge,xu2024triboelectrification,li2022spontaneous,sobolev2022charge}.

While some theoretical studies predict the significant friction and adhesion increase caused by electrostatic interactions \cite{wolloch2018,sun2023,persson2021general}, these theories typically ignore effects such as leakage current or charge trapping, which may be important in practice, for example, on oxide film covered surfaces. Experiments are therefore required, to explore how local electric fields and charge migration dynamics are coupled to nanoscale contact deformation, adhesion and friction.


In this paper, we experimentally demonstrate the asymmetric dependence of friction force on the bias voltage applied to a multi-asperity silicon-on-silicon interface. Specifically, we observe that the friction force increases with increasing positive bias voltage applied to the substrate silicon wafer relative to the grounded polysilicon ball, but remains unaffected under negative bias voltage with respect to the same grounding reference. The dependence of the friction force on the positive bias voltage is well captured by our electroadhesion model based on the contact map obtained from boundary element method (BEM) calculations. As the positive bias voltage increases, the electric field across the contact interfaces strengthens, resulting in the accumulation of more mobile charge carriers toward the interface, causing stronger electrostatic attraction and an associated increase in the friction force. In contrast, under negative bias, the friction force remains constant, which we attribute to the electronic properties of the polysilicon ball. We propose that mobile positive charge carriers become trapped within the grounded bulk polysilicon ball during their migration toward the interface in contact with a negatively biased wafer, weakening the Coulombic interactions between the charge pairs and thereby preventing electroadhesion.

\begin{figure}[h]
\includegraphics[width=0.4\textwidth]{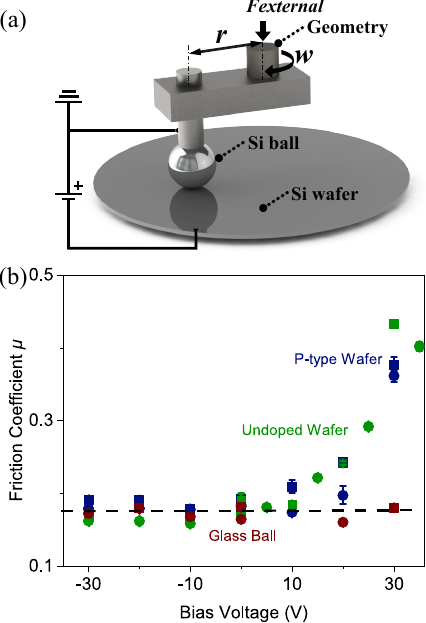}
\centering
\caption{ 
Experimental setup (a) and the measured bias voltage-dependent friction (b). (a) A 3-mm-diameter polysilicon ball, clamped within a brass ball holder, is brought into contact with a silicon wafer inside a dry chamber (RH=0.8\%). The ball holder is positioned at a distance  \( r \) from the rotation axis of the rheometer and is fixed to the rheometer's geometry through a connector. By imposing a constant angular velocity \( \omega \) from the rheometer, the rotating geometry moves the ball on the wafer at a corresponding sliding speed of \( \omega r \). The friction forces and normal loads were measured while maintaining the bias voltage \( V \) applied to the backside of the wafer and grounding the ball. 
(b) Friction coefficient, \(\mu\), as a function of the bias voltage \( V \). The blue, green and dark red data correspond to the contacts of polysilicon ball-on-P-type wafer, polysilicon ball-on-undoped wafer and glass ball-on-P-type wafer, respectively. To eliminate the influence of hysteresis (Fig.~S1), the applied bias voltage was first increased (solid squares) and then decreased (solid circles). The dashed line through the data points is drawn to guide the eye.
} 
\label{fig1}
\centering
\end{figure}

We measured silicon-on-silicon friction using a rheometer (DSR 502, Anton Paar) inside a dry chamber (RH=0.8\%), as shown in Fig.~\ref{fig1}(a) (see more details in Sec. A \cite{supplemental}). In the friction measurements, a cleaned polysilicon ball (with a native oxide layer) was brought into contact with an as-received monocrystalline silicon wafer (University Wafer) covered with a native oxide layer. The externally applied normal load was kept constant at 40 mN, corresponding to a Hertz contact radius of 8.5 $\mu$m. During sliding, both externally applied normal load ($F_{external}$) and dynamic friction force ($F_f$) at the dry contact interface were simultaneously measured by the rheometer. The ratio of these two forces gives the friction coefficient $\mu=F_f/F_{external}$. It is important to note that the measured friction coefficient may be affected by the adhesive force at the contact interface, which adds to the externally applied normal load and is balanced by the repulsive force generated at the contact points. The sliding speed imposed by the rheometer was fixed at 0.1 $\mu$m/s for all the measurements and the stable friction coefficient was reported after a few short sliding strokes. To investigate the dependence of friction on the applied bias voltage, the friction coefficient was measured while maintaining a target bias voltage ($V$) applied to the backside of the wafer using a power supply (Keysight E36232A), and grounding the polysilicon ball. The applied bias voltage was varied between -30 $V$ to 30 $V$. 

As shown in Fig.~\ref{fig1}(b), the dependence of the friction coefficient on the applied bias voltage at the silicon interfaces can be divided into two distinct regimes. In the positive bias voltage regime, the friction coefficient increases with increasing bias voltage. On the contrary, the friction coefficient remains unaffected under applied negative bias voltages. 

It is well established that a potential difference across a dielectric layer—such as the air gap and native oxide present at our contact interfaces—generates an electric field that drives the accumulation of opposite charge carriers on the opposing surfaces. This, in turn, results in electroadhesion at the interface \cite{sirin2019electroadhesion}, effectively contributing an additional normal load and thereby increasing friction force. By simplifying our biased contact interface as a charged plane-parallel capacitor, we get an electroadhesion force on the order of a mN with an applied bias voltage of $\sim 30\ V$, sufficient to affect the friction force (see more details in Sec.~B \cite{supplemental}). 


To refine the rough estimation and gain the local electroadhesion force distribution at the deformed contact interface, we propose a simple adhesion model to quantitatively link the electroadhesion force to the measured friction force. In this model, we attribute the interfacial adhesive force exclusively to the electroadhesion force generated in the non-contact (gap) area, where the local solid-air-solid gaps are treated as parallel capacitors. The electroadhesion force arising from solid–solid contact is neglected here and will be discussed later. We also neglect the contributions from capillary adhesion as dry experimental conditions suppresses the formation of capillary bridges across the interface \cite{peng2022}. Van der Waals interactions are not considered either, since the estimated van der Waals force is on the level of \(\sim 100\ \mu\)N based on the Derjaguin approximation with a function distance of 0.3 nm, insufficient to significantly affect the friction \cite{israelachvili2011intermolecular,hsia2022}. Consequently, the net force in the normal direction -the sum of the externally applied normal load ($\Vec{F}_{external}$) and the electroadhesion force ($\Vec{F}_{ea}$)- is balanced by the elastic repulsive force ($\Vec{F}_{elastic}$) at the interface, as expressed by:
\begin{equation}
\Vec{F}_{ea}+\Vec{F}_{external}+\Vec{F}_{elastic}=0 
\label{eq1}
\end{equation}

We assume a constant proportion between the friction force and the net normal force, or elastic repulsive force, in accordance with the principles of load-controlled friction,  as previously applied to similar silicon contact interfaces \cite{berman1998amontons,hsia2021rougher,peng2022}. This constant proportionality ratio ($\mu_0$) can be extracted from the friction force measured at the contact interface in the absence of electroadhesion force, i.e., under zero bias voltage ($\mu_{0}={F}_{f}({V=0})/{F}_{external}$). Therefore, we can calculate the net normal force, or the elastic repulsive force,  ${F}_{elastic}$, as: 
\begin{equation}
\begin{aligned}
\left|\Vec{F}_{elastic}\right| = \frac{F_f}{\mu_0} = \frac{\mu({V}) \times \left|\Vec{F}_{external}\right|}{\mu_0}
\label{eq2}
\end{aligned}
\end{equation}
where $\mu(V)$ is the measured friction coefficient under an applied bias voltage $V$.

To obtain the contact deformation for electroadhesion force estimation, we conduct elastic contact calculations using the boundary element method (BEM) \cite{BEM}. Using the elastic repulsive force from Eq.~\ref{eq2}, we obtain the contact deformation and the corresponding three-dimensional nanoscale deformed interface geometry. The local electric field ($E_i$) is then calculated as $E_i={\frac{V}{u_i+h_0}}$ \cite{persson2021general,persson2018dependency} (Fig. S3), where the $V$, $u_i$ and $h_0$ are the applied bias voltage, the local interfacial gap and the effective thickness of the native oxide layer respectively. The parameter $h_0$ is determined by the thickness of the native oxide layer ($d$ = 2.2 nm \cite{oxidethickness}) and the dielectric constant (${\kappa}$ = 3.9 \cite{itsumi1993influence}) of the native oxide layer, and is given by $h_0=2\times d / {\kappa} $. Finally, the total electroadhesion force (${F}_{sim-ea}$) is calculated as the sum of the locally experienced electroadhesion force (${F}_{sim-ea}^i$) along the gap area within the contact interface: 
\begin{equation}
\begin{aligned}
{F}_{sim-ea}&=\sum{F}_{sim-ea}^i=\sum \frac{1}{2}\epsilon_0 E^2_iA_i\\
&=\sum \frac{1}{2}\epsilon_0 ({\frac{V}{u_i+h_0}})^2 A_i
\label{eq3}
\end{aligned}
\end{equation}
where $A_i$ is the local gap area.

\begin{figure}[h]
    \centering
    \includegraphics[width=0.4\textwidth]{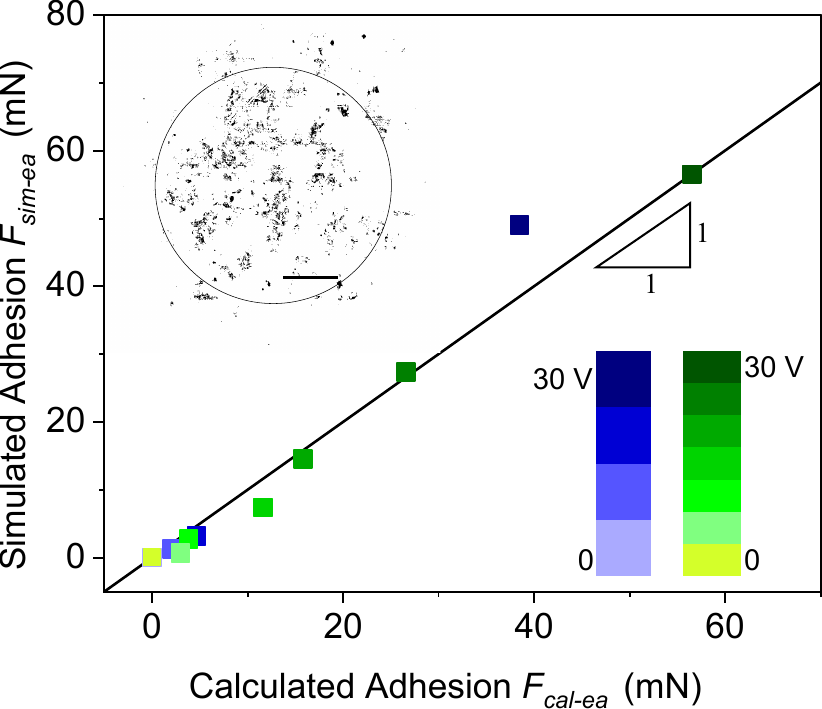}
    \caption{Adhesion forces calculated from friction experiments and simulations. The blue and green data correspond to the contacts between the polysilicon ball and the P-type wafer, and the polysilicon ball and the undoped wafer, respectively. The corresponding color bar indicates the applied bias voltage for each case. The upper inset shows the calculated contact map at the interface between a wafer biased at +30 V and a grounded polysilicon ball, with the white and black region corresponding to the solid-solid contact area and gap area,, respectively. The black circle denotes the Hertzian contact area. Scale bar, 5 $\mu$m.}
    \label{fig2}
\end{figure}

To verify our estimation of the electroadhesion force, we compare the simulated electroadhesion force (${F}_{sim-ea}$) with the calculated electroadhesion force (${F}_{cal-ea}$). The latter is derived from the measured friction coefficient at the biased interface using Eq. \ref{eq1} and Eq. \ref{eq2}: 
\begin{equation}
\begin{aligned}
{F}_{cal-ea} = \frac{\mu(V) \times {F}_{external}}{\mu_0} - {F}_{external}
\label{eq4}
\end{aligned}
\end{equation}

As shown in Fig.~\ref{fig2}, the electroadhesion forces obtained through the two methods show strong agreement, confirming the accuracy of our model and supporting our interpretation of the electroadhesion-enhanced friction in the positive bias voltage regime. We emphasize that the only adjustable parameter in our model is the thickness of the native oxide layer, for which we fit a value of 2.2 nm which is realistic for native oxide on silicon. Our modeled electroadhesion also agrees with the observed constant friction coefficient when a glass ball, instead of a polysilicon ball, is used in the biased friction measurements. In that case, the glass ball insulates charge carrier propagation and charge carriers can only accumulate at the interface between the glass ball and brass ball holder. In our model, this scenario corresponds to an $h_0$ on the same order as the glass ball diameter —millimeter-scale— resulting in a negligible electric field and thus minimal electroadhesion. Additionally, our atomic force microscopy  (AFM) adhesion measurements directly show an increasing adhesion force with increasing negative bias voltage at the contact between a grounded Si tip and negatively biased polysilicon ball (see more details in Sec.~A \cite{supplemental} and Fig.~S4), consistent with the increased friction coefficient measured between a grounded silicon wafer and a negatively biased polysilicon ball, further reinforcing our electroadhesion interpretation.

Although our model shows a strong agreement between the simulated and calculated electroadhesion forces in the positive bias voltage regime, it fails to predict the absence of electroadhesion forces under negative bias voltages. The measured friction coefficient under negative bias voltage remains consistent with that observed at zero bias voltage (Fig. \ref{fig1}(b)), indicating a negligible electroadhesion force. However, our model still predicts a nonzero electroadhesion force based on Eq.~\ref{eq3}. This discrepancy cannot be attributed to differences in wafer dopant types, as the friction shows similar response to bias voltages at the interface between a undoped silicon wafer and a grounded polysilicon ball. Tribocharging \cite{sayfidinov2018minimizing} appears negligible, given the stable friction force measured at zero bias voltage over the sliding distance. Furthermore, significant charge carrier depletion at the interface is unlikely due to the similar work functions of polycrystalline and monocrystalline silicon \cite{king1994electrical}.

\begin{figure}[!h]
\includegraphics[width=0.4\textwidth]{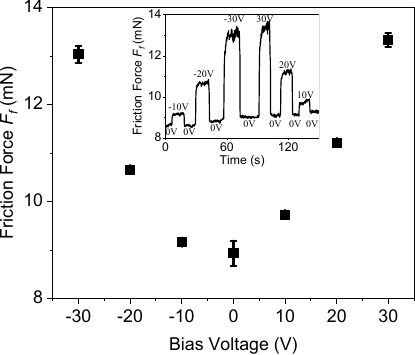}
\centering
\caption{
Average friction force measured as a function of the bias voltage applied on the substrate wafer at the wafer-on-wafer contact interface. The inset shows the real time variation of friction force over applied bias voltage during the sliding.
}
\label{fig3}
\centering
\end{figure}

To explore the adhesion discrepancy between our model and experiments in the negative bias voltage regime, we repeated the bias friction measurements at the contact between one piece of P-type wafer and the substrate P-type wafer, instead of the original polysilicon ball-on-wafer configuration. A glass pin, attached to the edge of a plate geometry (CP50-1/S, Anton Paar), was used to push the top grounded wafer (1 $\times$ 1 cm\textsuperscript{2}) to slide over the biased substrate wafer at a constant velocity of 0.1 $\mu$m/s under a fixed normal load of 23 mN, which includes both the body weight of the top wafer and the applied dead weight. During the sliding, the friction force was measured. Meanwhile, the bias voltage applied to the substrate wafer was varied between -30 $V$ and 30 $V$, in intervals of 10 $V$. Before each change to the target voltage, the substrate wafer was grounded (0 $V$) until a steady friction force was achieved. The measured friction force, shown in Fig. \ref{fig3}, increases with increasing the applied bias voltage and exhibits a symmetric response for both positive and negative bias voltages of equal magnitude. 

We therefore propose that the asymmetric adhesive response to opposite bias polarities (Fig. \ref{fig1}(b)) results from the electronic properties of the polysilicon ball. Specifically, mobile positive charge carriers may encounter increased difficulty traversing bulk polysilicon compared with the mobile negative charge carriers. Charge carrier traps within the bulk polysilicon may hinder the accumulation of positive carriers at the ball surface. This charge trap hypothesis is consistent with the friction hysteresis observed after bias removal at the grounded polysilicon ball-on-biased wafer contact (Fig.~S1): under the influence of the strong electric fields (Fig.~S3), charge carriers can get trapped at the silicon/oxide interface or within the oxide \cite{tzeng2006charge}. While polysilicon displays friction hysteresis that lasts for minutes (Fig.~S1), at the biased monocrystalline wafer-on-wafer interface such hysteresis is lacking (Fig.~3). Reported charge trap densities in polysilicon and oxide layers vary widely but can reach the level of $\sim$ 10\textsuperscript{13} cm\textsuperscript{-2} (equivalent to $\sim$ 16 mC/m\textsuperscript{2}) \cite{itsumi1993influence}. Although this is approximately an order of magnitude below our estimated surface charge density in the positively biased regime (Fig.~S3), volumetric traps in bulk polysilicon —especially at grain boundaries— can be sufficient to limit positive carriers transport over depths extending tens to hundreds of nanometers \cite{seto1975electrical}. Moreover, the charge density in our model might be overestimated, as it relies on the idealized assumption of an infinite parallel-plate capacitor with a uniform, fixed surface charge distribution. In our model, the charge traps suggest that the voltage drop only partly takes place at the contact interface, when the wafer is negatively biased relative to the grounded polysilicon ball. Such small voltage drop across the contact interface generates a weak electric field, resulting in minimal electroadhesion at the interfaces. When the polysilicon ball is replaced by a piece of wafer, the free mobility of positive and negative charge carriers within the bulk monocrystalline silicon wafer result in the same Coulombic interactions and thus electroadhesion force for both positive and negative bias voltages of equal magnitude.

To further support our interpretation, we directly measured the capacitance at the biased ball–on-wafer interface. Notably, an electric current is observed across the biased contact interface (Fig.~S5(a)). Moreover, the leakage current scales linearly with the calculated elastic repulsive force under various externally applied loads at a 30 V bias voltage applied to the wafer relative to grounded silicon ball (Fig.~S6). This proportionality supports the load-controlled friction assumption, as the current is expected to scale with the real contact area \cite{bowden1939area}, which is itself proportional to the elastic repulsive force \cite{peng2025decrease}. To characterize the current–voltage behavior, we derived Fowler–Nordheim plots (Fig.~S4(b)), from which tunneling energy barriers of approximately 1.05 eV and 1.69 eV were extracted for the P-type and undoped wafers, respectively, comparable to values reported for similar contact systems \cite{ando2000conducting}.

\begin{figure}[!h]
\includegraphics[width=0.4\textwidth]{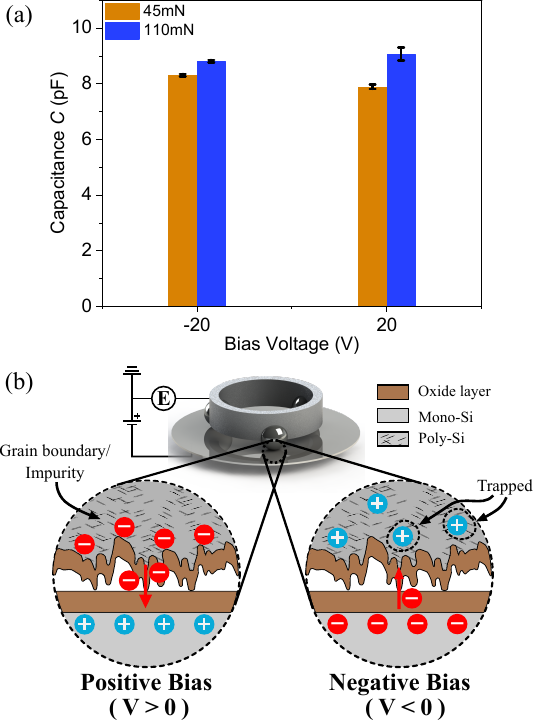}
\centering
\caption{
Capacitance measurement and the charge trapping mechanism. (a) The capacitance is measured under $\pm 20$ V bias voltage with increasing the externally applied loads from 45 mN to 110 mN. Capacitance is calculated as the ratio of absolute charge change to the applied bias voltage. (b) The upper panel shows a schematic of the experimental setup. Three grounded polysilicon balls clamped by an aluminum ring are put on top of a dry thermally oxidized silicon wafer and charges transferred to the polysilicon balls during charging are measured using a Keithley 617 electrometer. The bottom panel illustrates the polarity-dependent charge trapping mechanisms.
}
\label{fig5}
\centering
\end{figure}

To exclude the possibility that Joule heating from the tunneling current is responsible for the observed friction changes, we repeated the friction experiments at the contact between a positively biased wafer and a sandblasted silicon ball. The roughened surface of the sandblasted ball significantly increases the contact separation distance, thereby reducing the area where electroadhesion can occur. In this grounded ball-on-biased wafer contact, no electroadhesion effects were observed in the positive bias regime, even though a measurable leakage current was still present (Fig.~S7). This suggests that the polarity-dependent friction changes are not due to Joule heating. Nonetheless, our model remains valid even when such leakage current is ignored, as charge transfer through the contact interface does not eliminate the electroadhesion force generated among charge pairs in the gap regions. 
To suppress charge leakage in capacitance measurements, we grow a 50 nm oxide layer on the wafer using dry thermal oxidation.  Given the sensitivity of such small capacitance values ($\sim$ pF estimated from Fig.~S3) to subtle fluctuations, we focus on the capacitance increase with increasing applied load, thereby isolating the contribution of the expanding contact area from all other sources of fluctuations (see details in Fig.~S8). As shown in Fig.~\ref{fig5}(a), the capacitance increases \(14.83\% \pm 3.13\%\) under the positive bias voltage, compared with a \(6.04\% \pm 1.84\%\) capacitance increase under negative voltages of the same magnitude with the load increasing from 45 mN to 110 mN. This asymmetric increase is consistent with our charge carrier mobility hypothesis shown in Fig.~\ref{fig5}(b): positive charge carriers with reduced mobility in bulk polysilicon effectively increases the dielectric thickness, thereby lowering the interfacial capacitance between the negatively biased wafer and the grounded polysilicon ball.

In conclusion, we have demonstrated the influence of the electroadhesion on friction at multi-asperity silicon-on-silicon interfaces by applying  bias voltages across the interfaces. We observe an asymmetric dependence of friction force on the bias voltage applied to the substrate wafer: while a positive bias voltage increases friction force, a negative bias voltage has no significant effect on friction force. Using a simple electrostatic adhesion model, we capture the increase in friction force and attribute it to the electroadhesion, which acts as an additional normal load arising from the non-contact area at the interfaces. The good agreement between the model and experiments validates the theoretical electrostatic framework, extending the electroadhesion predictions to interfaces with nanoscale roughness. Remarkably, the electroadhesion can reach the order of $\sim$10 mN over a limited area as small as $\sim$100 \(\mu\)m\textsuperscript{2} at high surface charge density around 0.1 C/m\textsuperscript{2}. Conversely, the electroadhesive effect could also be important at lower charge density but for larger adhesive areas, for example at tribocharged interfaces (10\textsuperscript{-5}$\sim$ 1 mC/m\textsuperscript{2}) \cite{pan2019fundamental,baytekin2011mosaic,liao2024charge,grosjean2023single}. In contrast, under negative bias, charge trapping effectively immobilizes mobile carriers and screens electrostatic attraction, suppressing electroadhesion and its impact on friction. This polarity-dependent susceptibility to electroadhesion in polysilicon offers a strategy for designing interfaces with tunable adhesion and friction, enabling precise control where these effects are either desired or need to be minimized. All in all, our result highlights electroadhesion at stiff interfaces with nanoscale roughness, thereby opening up new avenues for friction control, for instance in precision positioning in the semiconductor industry.

\textit{Acknowledgments}---We thank Tijs van Roon for discussion. We thank Bob Drent and Kasper van Nieuwland for sample preparation.

%


\end{document}


\preprint{APS/123-QED}

\title{\mbox Supplemental Material for: Polarity-Dependent Electroadhesion at Nanoscale Rough Silicon Interfaces}

\author{Liang Peng$^{1}$\footnotemark[1]{*}}
\author{Stefan Kooij$^{1}$} 
\author{H. Tun\c{c} \c{C}ift\c{c}i$^{2}$}  
\author{Daniel Bonn$^{1}$} 
\author{Bart Weber$^{1,2}$}  
\affiliation{$^{1}$ Van der Waals-Zeeman Institute, Institute of Physics, University of Amsterdam, Science Park 904, 1098 XH Amsterdam, The Netherlands \\
$^{2}$ Advanced Research Center for Nanolithography (ARCNL), Science Park 106, 1098 XG Amsterdam, The Netherlands}

\maketitle
\onecolumngrid 

\subsection*{Section A: Experimental and Numerical Methods}

\vspace{1em}
\noindent\textbf{A1. Friction Measurement}

The friction was measured using a rheometer (DSR 502, Anton Paar) inside a customized humidity controlled chamber. Before friction measurements, the undoped polycrystalline silicon (polysilicon) balls (Goodfellow) were sonicated in ethanol and then Milli-Q water, followed by nitrogen flow drying. We used the boron-doped (P-type) and undoped silicon wafers in different measurements to investigate the dopant impacts on bias friction. Before the contact was formed, the surfaces were equilibrated under the dry nitrogen environment inside the chamber for one hour, during which the relative humidity was stabilized at 0.8\% by continuously flowing dry nitrogen. To minimize the influence of wear in the friction measurements, the sliding distance in each stroke was kept at 2 $\mu$m, and each stroke was measured on a previously untouched and fresh area on the silicon wafer. Before each change in the applied bias voltage, several sliding strokes were performed while grounding both the silicon wafer and the polysilicon ball until the friction coefficient returned to the value measured at the initial unbiased contact interface, thereby eliminating any hysteresis effects (Fig. \ref{figs1}).

\vspace{1em}
\noindent\textbf{A2. Contact Calculation}

We used BEM in the contact calculation, in which the contact between a silicon ball and a silicon wafer is modeled as a single rough surface in contact with a rigid flat surface, with the roughness from both silicon surfaces mapped onto the single rough surface. In the BEM calculation, in-plane contact deformation is numerically solved based on the Boussinesq solution for an elastic half-space under an applied load, assuming the local contact pressure within the real contact area is positive and below the material's hardness. When the local contact pressure exceeds the hardness, the model captures the deformation as perfectly plastic behavior.  The inputs for our contact calculations included the elastic repulsive force at the interface, the surface topography of the polysilicon ball and the silicon wafer (Fig. \ref{figs2}), and the mechanical properties of silicon: Young's modulus: \(E = 130 \pm 3 \, \text{GPa}\), Poisson's ratio: \(\nu = 0.26\). In the contact calculation, we assume that the locally experienced electroadhesion force influences the contact deformation, or the average interfacial gap, in the same way as the externally applied normal load, consistent with the framework of a DMT-type contact model \cite{ciavarella2020simplified,derjaguin1975effect}.

\vspace{1em}
\noindent\textbf{A3. AFM Adhesion Measurement}

We conducted atomic force microscopy (AFM, Dimension Icon, Bruker) adhesion measurements on a biased polysilicon ball, in which a grounded sharp n-type antimony doped Si tip (radius $5\sim12$ nm, PFQNE-AL, Bruker) was brought toward a negatively biased polysilicon ball and retracted afterwards, with the interaction force ($F$) and displacement ($D$) recorded. The adhesion force, or the absolute value of the pull-off force, averaged over 4096 individual pull-off measurements is then extracted from the typical $F-D$ curve and presented in Fig. \ref{figs4}. 

\vspace{1em}
\noindent\textbf{A3. Capacitance Measurement}

We used the Keithley 617 electrometer to measure the capacitance at the interface between a polysilicon ball and an oxidized wafer. The wafer oxidation was performed at 500 sccm O\textsubscript{2}, 1 atm, and 1100 °C for 442 seconds using Rapid Thermal Annealer (Sitesa addax RM6). The thermally grown oxide thickness was measured using a Filmetrics F20 UVX. During the capacitance measurement, the contact system was initially discharged to reset its charge state for 10 s, followed by a 10-s charging, and then discharged again for 10 s. The charge transferred to the polysilicon balls during the discharge–charge–discharge cycle was recorded by the electrometer, and the capacitance was determined as the ratio of the absolute charge variation to the applied bias voltage

\subsection*{Section B: Experimental and Simulation Results}
\renewcommand{\thefigure}{S\arabic{figure}}

\vspace{1em}
\noindent\textbf{B1. Capacitance Estimation}

To roughly estimate the electroadhesion force, we simplify the biased contact interface as a charged plane-parallel capacitor, where the silicon wafer and polysilicon ball are modeled as parallel plates with a dielectric film (air) in between. The capacitor's area is approximated by the Hertzian contact area ($\sim$ 270 \textmu m\textsuperscript{2}), and the dielectric thickness is taken to be the magnitude of the RMS (root-mean-square) roughness of the polysilicon ball ($\sim$ 40 nm, see Fig. S3). This simplified model suggests an electroadhesion force on the order of a mN with an applied bias voltage of $\sim 30\ V$, comparable to the external load and thus sufficient to affect the friction force. Additionally, the corresponding surface charge density is approximately \(\sim 10^{17} \) m\textsuperscript{-2} (equivalent to  0.016 C/m\textsuperscript{-2}). Considering the typical doping levels (\(10^{22}\sim 10^{26} \) m\textsuperscript{-3}) observed in semiconductor devices \cite{arora1982,wang2019,gharbi2011} and a wafer thickness of \(\sim 500 \, \mu\text{m} \), the wafer can supply sufficient mobile charge carriers to achieve a surface density at the order of \(10^{18}\sim 10^{22} \) m\textsuperscript{-2}.

\vspace{1em}
\noindent\textbf{B2. Friction, Electroadhesion, Electric Current, and Capacitance Characteristics}

\begin{figure*}[th]
\includegraphics[width=0.8\textwidth]{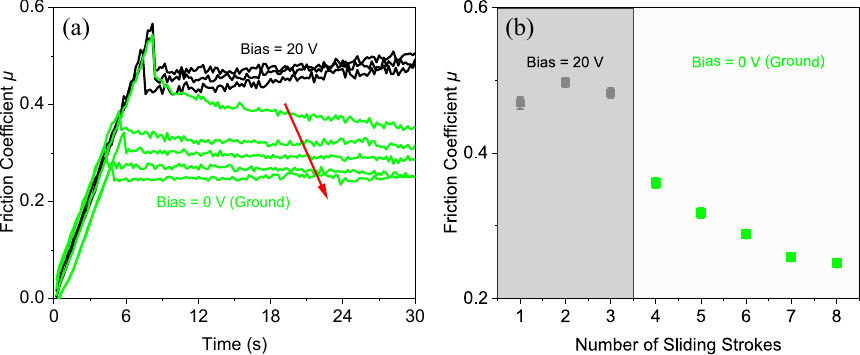}
\centering
\caption{Bias voltage-induced hysteresis effect on friction. (a) The friction coefficient is measured with the wafer initially biased at 20 V and then grounded, while the polysilicon ball remains grounded throughout. The black and green curves represent the friction measured while maintaining the applied bias voltage and after grounding the wafer, respectively. All sliding strokes are recorded continuously. The arrow indicates the decrease in friction coefficient upon switching the bias state to the grounded state. (b) The measured friction coefficient as a function of sliding strokes, corresponding to (a).}
\label{figs1}
\centering
\end{figure*}

\begin{figure*}[th]
\includegraphics[width=0.8\textwidth]{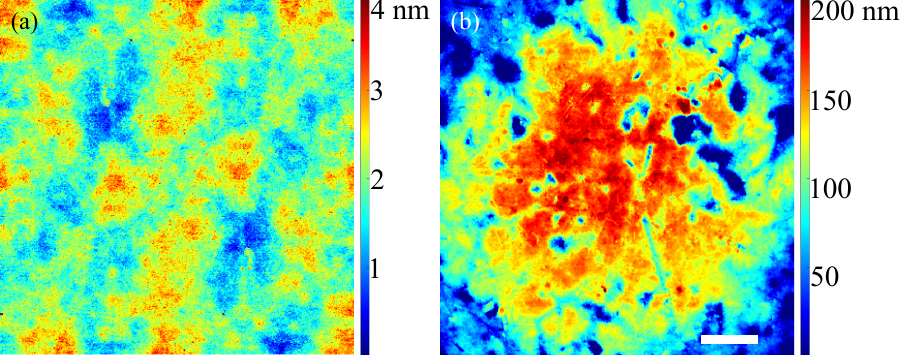}
\centering
\caption{
AFM topography of silicon wafer (a) and polysilicon ball (b). Measurements were taken using tapping mode in AFM (Dimension Icon, Bruker) with Si tips (RTESPA-300, Bruker) over an area of 31.13 × 31.13 $\mu\text{m}^2$. The RMS (root-mean-squared) roughness of the surfaces were 0.5 nm (wafer) and 40.5 nm (ball), respectively. Scale bar, 5 $\mu$m.
}
\label{figs2}
\centering
\end{figure*}

\begin{figure*}[th]
\includegraphics[width=0.8\textwidth]{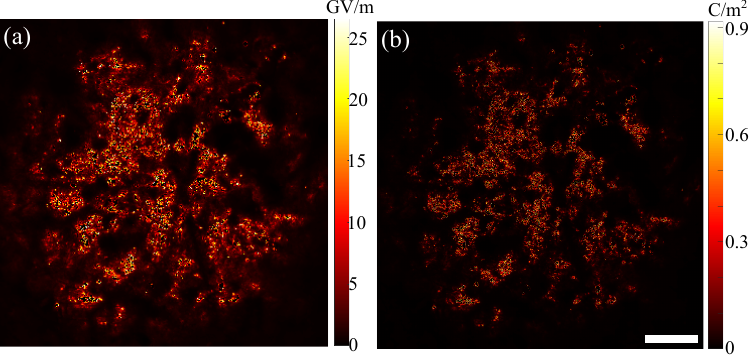}
\centering
\caption{In-plane local electric field, $E_i$, distribution (a) and its corresponding charge density, \(\sigma_i\), distribution (b) at the interface between the biased wafer (30 V) and grounded ball under an externally applied load of 40 mN. The charge density is estimated using Gauss’s law as  \( \sigma_i = \kappa \varepsilon_0 E_i \). Scale bar, 5 $\mu$m.}
\label{figs3}
\centering
\end{figure*}

\begin{figure*}[th]
\includegraphics[width=0.4\textwidth]{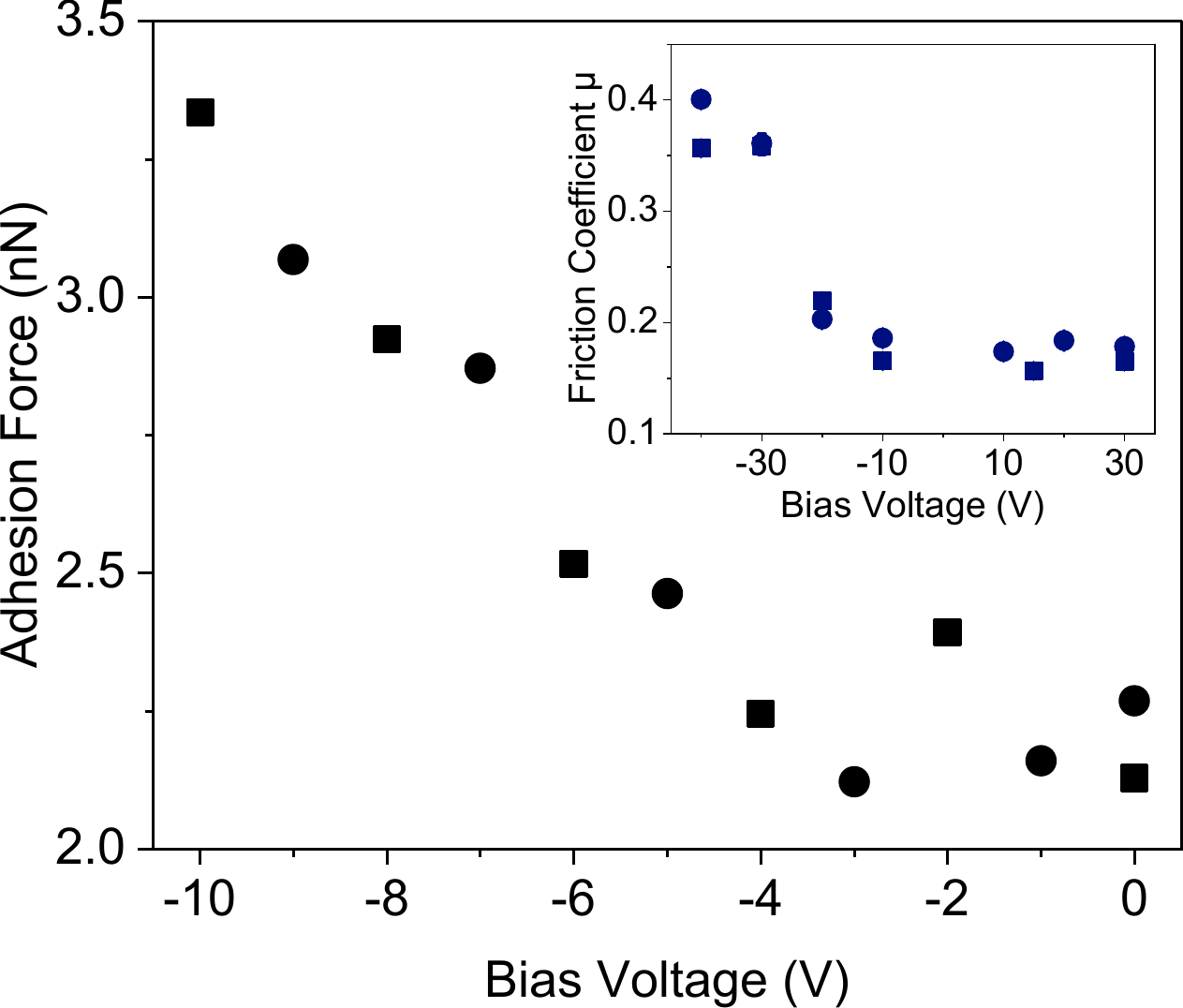}
\centering
\caption{
Adhesion force as a function of bias voltage applied to a polysilicon ball. Each data point represents the average of 4096 individual pull-off measurements taken from different spots within an area of 500 $\times$ 31 nm\textsuperscript{2} on the apex of the biased polysilicon ball. It should be noted that, although the applied bias voltage here is limited within 10 $V$ due to the setup configuration, a clear trend of increasing adhesion force with increasing negative bias is observed. This increasing trend agrees well with the increased friction coefficient measured at the interface between a grounded silicon wafer and a negatively biased polysilicon ball as shown in the inset. In both experiments, solid squares and circles correspond to measurements during increasing and decreasing bias, respectively.}
\label{figs4}
\centering
\end{figure*}

\begin{figure*}[th]
\includegraphics[width=0.8\textwidth]{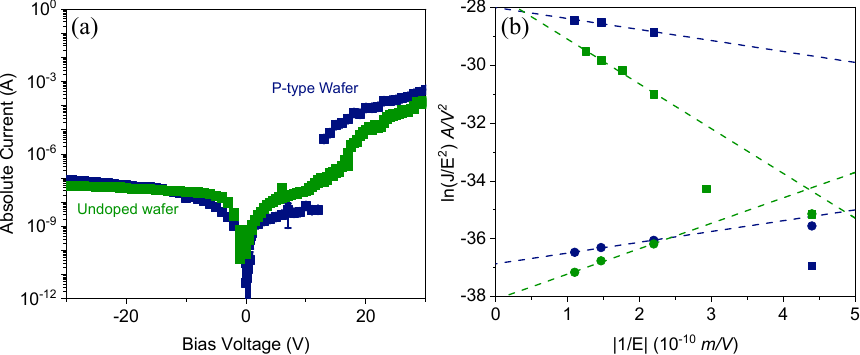}
\centering
\caption{
Current–voltage measurement and corresponding Fowler–Nordheim plot (a) Absolute electric current flowing through the contact interface as a function of bias voltage applied to the wafer relative to the grounded polysilicon ball. The electric current is measured using Keithley DMM6500. The blue and green data correspond to the contacts of polysilicon ball-on-P-type wafer and polysilicon
ball-on-undoped wafer, respectively. (b) The corresponding Fowler–Nordheim plots for the measured current–voltage data. The natural logarithm of the absolute $J/E^2$ is plotted against the $|1/E|$, where $J$ is the current density, calculated as the ratio of the measured the current to the modeled real contact area, and $E$ is the electric field,  defined as the applied bias voltage divided by the total thickness of the native oxide layer (4.4 nm). Squares and circles correspond to data obtained under positive and negative bias voltages, respectively. Dashed lines represent linear fits to the data.
}
\label{figs5}
\centering
\end{figure*}

\begin{figure*}[th]
\includegraphics[width=0.8\textwidth]{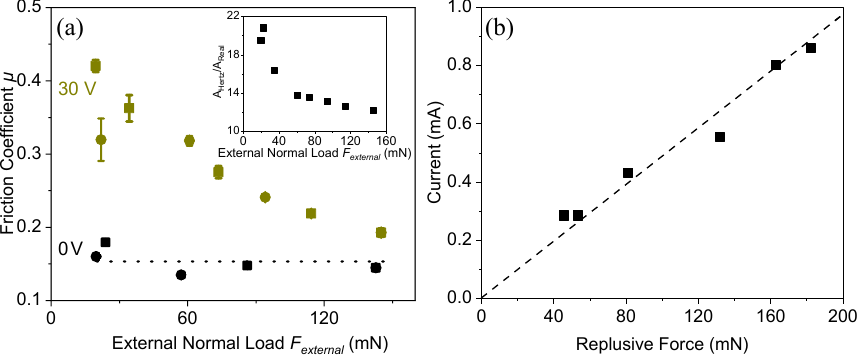}
\centering
\caption{External load-weakening friction and the corresponding repulsive force-strengthening leakage current at the interface between a biased P-type wafer and a grounded polysilicon ball. (a) Compared to the constant friction coefficient measured at unbiased interfaces, the friction coefficient decreases with increasing externally applied normal load. The inset shows the the ratio of the Hertzian contact area to the corresponding real contact area (\(A_{Hertz}/A_{Real}\)) under various external loads. The decreasing trend of \(A_{Hertz}/A_{Real}\) aligns well with the weakening friction over the external load as the relative importance of the gap area, or the electroadhesion area, shrinks with increasing load. (b) The measured leakage current increases with the elastic repulsive force (calculated from Eq. 4 in the main text) at a constant, 30 V bias voltage applied to the P-type wafer. This increase in leakage current arises from the growing real contact area as the external load increases.
}
\label{figs6}
\centering
\end{figure*}

\begin{figure*}[th]
\includegraphics[width=0.5\textwidth]{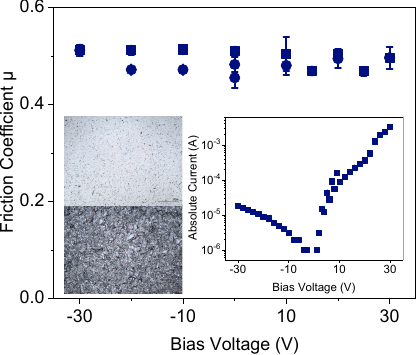}
\centering
\caption{
Bias-independent friction measured from the interface between a biased P-type wafer and a grounded sandblasted rough polysilicon ball. The friction coefficient was measured while biasing the wafer and grounding the polysilicon ball at a constant applied external load of 40 mN and 0.8\% RH. The applied bias voltage was first increased (solid squares) and then decreased (solid circles) to exclude any hysteresis effects. The left inset shows the contrast in topography between the original smooth polysilicon ball (upper) and the sandblasted rough polysilicon ball (lower) measured from a 3D confocal laser scanning microscope (Keyence VK-X1000). The scanned topography size is 283 $\times$ 212 $\mu$m\textsuperscript{2} with the RMS (root-mean-square) roughness around 64 nm (smooth) and 1439 nm (rough), respectively. Scale bar, 50$\mu$m. The right inset shows the measured electric current flowing across the interface as a function of bias voltage.}
\label{figs7}
\centering
\end{figure*}

\begin{figure*}[th]
\includegraphics[width=0.5\textwidth]{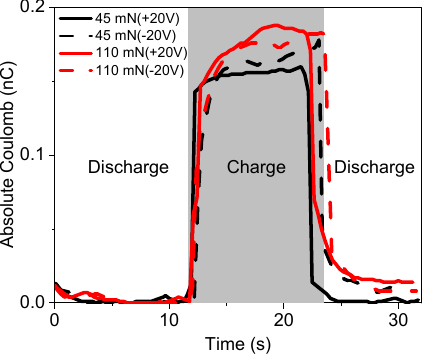}
\centering
\caption{
Capacitance measurement at the interface between a polysilicon ball and an oxidized wafer. The curves show the real-time change of absolute charge during the discharge–charge–discharge cycle under an applied wafer bias of 20 V, measured with a Keithley 617 electrometer at varying normal loads.
}
\label{figs8}
\centering
\end{figure*}

\clearpage
\newpage

%
